# Heywood cases in unidimensional factor models and item response models for binary data

Selena Wang, The Ohio State University

Paul De Boeck, The Ohio State University

Marcel Yotebieng, Albert Einstein College of Medicine

Author's Note

The data for the application are from a study funded by grant R01HD075171 from NIHCD.

Corresponding author: Selena Wang, Quantitative Psychology, Ohio State University,

Columbus, OH 43210. E-mail: wang.10171@osu.edu.

Abstract

Heywood cases are known from linear factor analysis literature as variables with communalities larger than 1.00, and in present day factor models, the problem also shows in negative residual variances. For binary data, ordinal factor models can be applied with either delta parameterization or theta parametrization. The former is more common than the latter and can yield Heywood cases when limited information estimation is used. The same problem shows up as nonconvergence cases in theta parameterized factor models and as extremely large discriminations in item response theory (IRT) models. In this study, we explain why the same problem appears in different forms depending on the method of analysis. We first discuss this issue using equations and then illustrate our conclusions using a small simulation study, where all three methods, delta and theta parameterized ordinal factor models (with estimation based on polychoric correlations) and an IRT model (with full information estimation), are used to analyze the same datasets. We also compared the performances of the WLS, WLSMV, and ULS estimators for the ordinal factor models. Finally, we analyze real data with the same three approaches. The results of the simulation study and the analysis of real data confirm the theoretical conclusions.

# Heywood cases in unidimensional factor models and item response models for binary data

Heywood cases are variables with a proportion explained variance (i.e., the communality) larger than 1.00 in factor analytic models, an anomaly for evident reasons. Harman and Fukuda (1966) named this problem after Heywood (Heywood, 1931), its first describer, and since then "Heywood cases" has become a common term in applied and methodological research. In factor models of today, with covariance matrix-based estimation, Heywood cases are identified by the appearance of negative residual variances. Heywood cases are far from exceptional (Chen, Bollen, Paxton, Curran, & Kirby, 2001; Jöreskog & Lawley, 1968; Martin & McDonald, 1975), and when they occur, the corresponding factor solutions are considered improper solutions (e.g., Jöreskog & Lawley, 1968). Whereas Heywood cases are not uncommon for factor models, linear or ordinal factor models, they do not occur in item response (IRT) models. In addition to factor models, IRT models constitute a prominent family of latent variable models for categorical variables, and the two are known to be connected (Kamata & Bauer, 2008; Takane & De Leeuw, 1987). It is therefore natural to wonder why factor model estimation runs into Heywood cases, but IRT model estimation does not.

For ordered-category variables, including binary variables, IRT models as well as factor models can be used. The two-parameter logistic (2PL) unidimensional and multidimensional IRT models are common models for binary data. The probit version of these IRT models can be estimated using ordinal factor models based on polychoric correlations (i.e., tetrachoric correlations for binary variables), the same as the ordinal factor models.

There are roughly speaking three major differences between factor models and IRT models:

(1) Most factor models are linear models for continuous and normally distributed variables. For ordered-category variables, including binary variables, one can rely on ordinal factor models. IRT models are always for categorical variables.

(2) For ordinal factor models, two categories of parameterizations are available[1] (Asparouhov & Muthén, 2020; Paek, Cui, Gübes, & Yang, 2018). Delta parameterization is the default and implies that the total variance of the hypothetical and normally distributed variable underlying the observed categorical variable (i.e., the latent response variable) is set to 1.00, which is in correspondence with the assumption of polychoric correlations. Theta parameterization is an alternative in which not the total variance of the latent response variable, but the residual variance instead is set to 1.00, and the explained variance and the factor loading are expressed relative to the fixed residual variance. An IRT model uses either a probit link or a logit link for the probability of responses. An IRT model can be formulated as a latent response model with a fixed error variance (e.g., Takane & de Leeuw, 1987) using either the standard normal variance of 1.00 (probit IRT models) or the standard logistic variance of 3.29 (logistic IRT models). The latter is the most common choice, but the parameterization of the former (i.e., the probit IRT model) corresponds to that of the theta parameterized ordinal factor model because both have the fixed error variance of 1.00.

(3) Estimation of factor models is commonly based on the variance-covariance matrix. More specifically for ordinal factor models, it is based on the polychoric correlation matrix (tetrachoric correlations for binary data). For linear factor models, the variance-covariance matrix is a sufficient statistic; for ordinal factor models, the polychoric correlation matrix does not entail all information in the data, and therefore the estimation based on polychoric correlations is called a limited information approach. Although a full information approach with maximum likelihood is available for ordinal factor models, it is not commonly used. Meanwhile, for IRT models, estimation is always based on the full data, instead of pairwise relationships between variables. For an overview and discussion of these estimation methods, see Bolt (2005), among others.

---

[1] In this paper, the latent factor is assumed to be a standard normal distribution for both the delta and the theta parameterizations.

The three approaches available for ordered-category data, which include binary data, are delta and theta parameterized ordinal factor models and IRT models, and only one of these three can lead to negative residual variances. Negative residual variances cannot occur in theta parameterized factor models, as they can in delta parameterized factor models, because residual variances in theta parameterized factor models are fixed to 1.00. Similarly, negative error variances cannot occur in IRT models because the residual variances in IRT models are also fixed, either to 1.00 or to the logistic error variance of 3.29. Therefore, it is natural to wonder what happens to theta parameterized ordinal factor models and to IRT models if negative residual variances are encountered in delta parameterized ordinal factor models.

Aim and overview

The aim of our study is to describe and explain anomalies in the results of the three approaches if a Heywood case is expected for the data under consideration. We focus on binary data, but the same principles apply to polytomous ordered-category data fitted to an ordinal factor model or to a partial-credit or graded-response IRT model.

In the next section, we describe the different approaches. We first introduce delta parameterization and then answer the question of what would happen if the same data showing a Heywood case in a delta parameterized ordinal factor model were fitted to a theta parameterized factor model or an IRT model. In the sections after that, we illustrate our conclusions with a small simulation study and a real dataset; the data in both illustrative sections are binary. Finally, we discuss implications of our findings.

## Three approaches

**Delta parameterization**

The observed values of ordered-category variables in an ordinal factor model are assumed to stem from underlying continuous variables (latent response variables) with a normal distribution and thresholds for discretization (into ordered-category response options). For binary variables, just one threshold is needed to differentiate between the observed values of 0 and 1. For identification reasons, a parameterization choice needs to be made for the latent response variables (denoted as $V$s). Delta parameterization means that the total variance of the latent response variable is set to 1.00. A Heywood case means that more than 100% of the variable's variance is explained by the (latent) factor(s), which implies that the residual variance is negative. For a model with one factor, it means that the (standardized) loading is larger than 1.00.

The equation for the variance of the latent response variable with delta parameterization, $V_{\delta i}$, is:

$$\sigma^2_{V\delta i} = \lambda^2_{\delta i} + \sigma^2_{\delta \varepsilon i}, \tag{1}$$

with $\sigma^2_{\delta V i} = 1$ as the variance of the latent response variable $i$ in the delta parameterized model, $\lambda^2_{\delta i}$ as its squared loading, and $\sigma^2_{\delta \varepsilon i}$ as its residual variance. To avoid having the total variance $\sigma^2_{V\delta i}$ exceeding 1.00, the model estimates the residual variance as negative if $\lambda_{\delta i} > 1$.

Theta parameterization

For theta parameterization, the ordered-category variables in an ordinal factor model are also assumed to stem from the continuous latent response variables, but for this type of parameterization, the residual variance, instead of the total variance, is set to 1.00. The equation for the variance of the latent response variable with theta parameterization, $V_{\theta i}$, is:

$$\sigma^2_{V\theta i} = \lambda^2_{\theta i} + \sigma^2_{\theta \varepsilon i}, \tag{2}$$

with $\sigma^2_{V\theta i}$ as the variance of variable $i$ in the theta parameterized model, $\lambda^2_{\theta i}$ as its squared loading, and $\sigma^2_{\theta\varepsilon i} = 1.00$ as its residual variance. The theta parameterized factor loading, $\lambda_{\theta i}$ can be translated into the delta parameterized factor loading, $\lambda_{\delta i}$ as follows:

$$\lambda_{\delta i} = \sqrt{\lambda^2_{\theta i}/(1 + \lambda^2_{\theta i})}. \tag{3}$$

Under the square root is the ratio of explained variance ($\lambda^2_{\theta i}$, given that the factor variance is 1.00) versus the total variance ($1 + \lambda^2_{\theta i}$). It follows that:

$$\lambda_{\theta i} = \sqrt{\lambda^2_{\delta i}/(1 - \lambda^2_{\delta i})}. \tag{4}$$

Equation (3) shows that a theta loading when translated into a delta loading yields a delta loading smaller than 1.00 (with a limit value of 1.00), and Equation (4) shows that the translation of a delta loading into a theta loading is problematic if $\lambda^2_{\delta i} \geq 1$. A Heywood case with delta parameterization (i.e, $\lambda_{\delta i} > 1$) would lead to a negative value under the square root in Equation (4). For a factor to explain 100% of the variance of a variable (the boundary condition of Heywood cases), the delta parameterized factor loading, $\lambda_{\delta i}$ would need to be 1.00, and the theta parameterized loading, $\lambda_{\theta i}$ would be infinitely large because the proportion explained variance is $\lambda^2_{\theta i}/(1 + \lambda^2_{\theta i})$. When $\lambda_{\delta i} = 1.00$, the proportion explained variance is 1.00, making $\lambda_{\theta i}$ infinitely large (see Equation (4)). Iterations on the way to extremely high proportions of explained variance would necessarily lead to convergence issues because even large increases in an already large theta loading would translate into extremely small and barely noticeable increases in proportions of explained variance as well as into barely noticeable decreases of the discrepancy between the polychoric correlations and the model-based correlations. This is equivalent with an optimization process that reaches a plateau and cannot find the highest point on the plateau, or equivalently, the lowest point in a flat valley. Therefore, when applying the same data to factor models with delta parameterization and with theta parameterization, we expect a Heywood case in the delta parameterized factor model to cooccur with (1) a nonconvergence case in the theta parameterized factor model, and (2) extremely high theta parameterized loadings when the iterations are stopped without convergence.

We believe that some of the nonconvergence cases encountered in Paek et al. (2018) in theta parameterized factor models may show as Heywood cases if the same data were fitted to delta parameterized models instead. Paek et al. simulated binary IRT data and analyzed these data with (1) theta parameterized factor models using three different limited information least squares estimators in Mplus (Muthen & Muthen, 1998-2017): fully weighted least squares (WLS), diagonally weighted least squares (WLSMV), and unweighted least squares (ULS), and (2) logistic and probit IRT models using full information Maximum Likelihood (ML) estimation. The WLS estimator is also known as the ADF approach (fully weighted least squares with no distributional assumption (Browne, 1984), and the diagonally weighted least squares is implemented if WLSMV is selected as the estimator in Mplus and in the cfa ( ) function from lavaan (Rosseel, 2012). Paek et al. (2018) aimed to compare parameter recovery and standard errors, but the authors also found a substantial nonconvergence rate with WLS, much more than with WLSMV and ULS, while nonconvergence did not occur with the IRT approach. Although Paek et al. did not use delta parameterization and did not consider Heywood cases, we suspect that some of the datasets showing convergence issues with theta parameterization would show Heywood cases in an analysis with delta parameterization.

IRT approach

To estimate IRT models, a marginal maximum likelihood (MML) approach based on full data information is commonly used, rather than limited information statistics such as the polychoric correlations. Because the optimization is not formulated in terms of the polychoric correlation matrix, the previous convergence issues, stemming from extremely small improvements in recovering the polychoric correlations, do not occur. This explains why Paek et al. (2018) never run into convergence issues using full information IRT based estimation despite that IRT models, the same as theta parameterized factor models, assume fixed residual variances.

When data showing Heywood cases in delta parameterized factor models are fitted to IRT models, we expect extremely high discrimination values (i.e., the "loadings" in an IRT model). An extremely high discrimination implies an extremely high proportion of explained variance. For Heywood cases, the proportion explained variance (of the latent response variable) exceeds 1.00, which is not possible when the residual variance is fixed, as is the case in theta parameterized factor models and in IRT models. However, we do not expect the same convergence issues in an IRT model as in a theta parameterized factor model with a limited information approach because the pairwise relationships between variables are not the target of IRT optimization. The relationships between variables are only one aspect of the full information used for estimating IRT models.

## Simulation-based Illustration

**Method**

To illustrate the differences of the three approaches (delta and theta parameterized factor models with limited information estimation and IRT models with full information estimation), we conducted a small simulation study with just 100 datasets, each with a sample size of 200, N=200. Only a small simulation study is needed because the purpose is to illustrate the differences, not to find an answer to a research question. The data were generated in two steps: obtaining normally distributed continuous data in the first step and dichotomizing the continuous data into binary data in the second step.

First, we generated 100 datasets, each with four normally distributed continuous variables with means of 0.00, variances of 1.00 and a covariance matrix (in this case, a correlation matrix due to the unit variance) as shown in Table 1. The data were generated with the mvrnorm () function from the MASS package in R (Venables & Ripley, 2002) without enforcing the estimated covariance matrix to be identical to the theoretical (true) covariance matrix (accomplished by using the option empirical=FALSE in the mvrnorm () function). The theoretical covariance matrix in Table 1 was chosen in such a way that there would be a substantial probability for the generated data to show Heywood cases after dichotomization. An important condition we specified for the theoretical covariance matrix was that continuous data generated with the mvrnorm () function with the empirical=TRUE option would lead to a Heywood case when a one-dimensional linear factor model is fitted to the data using delta parameterization. Using the default ML estimator in the cfa () function from lavaan 0.6-7 (Rosseel, 2012), the residual variance of the 3rd variable (Y3) had indeed a negative value, which gave us confidence in the proposed covariance matrix.

Next, for the 100 continuous datasets generated with the empirical=FALSE option, a threshold of 0.00 was used to dichotomize the data, the same threshold for all variables in all datasets. In this way, 100 binary datasets were obtained.

We expect (1) a substantial proportion of Heywood cases out of all fitted ordinal factor models with delta parameterization, (2) a substantial proportion of nonconvergent cases out of all fitted ordinal factor models with theta parameterization and (3) a substantial proportion of IRT models showing extremely large discrimination parameter estimates. Based on the theoretical covariance matrix, we expect variable Y3 to be the problematic variable for most datasets showing: a negative residual variance with delta parameterization, a extremely high loading when the iterations are stopped due to nonconvergence with theta parameterization, and an outlying discrimination parameter value with IRT.

Table 1

Theoretical covariance matrix to generate continuous data for the illustrative simulation study

|  | Y1 | Y2 | Y3 | Y4 |
| --- | --- | --- | --- | --- |
| Y1 | 1.00 |      |      |      |
| Y2 | 0.10 | 1.00 |      |      |
| Y3 | 0.48 | 0.76 | 1.00 |      |
| Y4 | 0.48 | 0.48 | 0.48 | 1.00 |

The 100 dichotomized datasets were analyzed with the cfa () function of lavaan 0.6-7 using the limited information estimators including WLS (i.e., ADF), WLSMV (diagonally weighted least squares) and ULS (unweighted least squares), and each estimator was used in both delta and theta parameterized ordinal factor models. The data were also fitted to an IRT model with the grm() function from the ltm package (Rizopoulos, 2006). The frequency of Heywood cases and the range of the resulting negative residual variances were determined for each estimator with delta parameterization. For the WLSMV estimator, which is the recommended estimator, a more detailed description of the negative values is given. The frequency of nonconvergence cases are reported per estimator with theta parameterization, and the range of extreme loadings when the iterations are stopped under nonconvergence are determined. Finally, for the IRT approach, the frequency of nonconvergence cases and an overview of estimated discriminations are reported.

**Results**

*Delta parameterization*. Using WLSMV and ULS as estimators, the fitted ordinal factor models with delta parameterization never showed convergence issues, but they did show Heywood cases for 34 and 33 out of the 100 datasets, respectively. The program reported warning messages indicating that negative variances were observed. In contrast, using WLS as estimators, the fitted ordinal factor models with delta parameterization reported Heywood cases in 51 out of the 100 datasets and failed to converge in 4 out of the 100 generated datasets. When the models failed to converge, extremely negative values of the residual variances (-2626745, -694577, -4376004 and -1309322) were observed in the output. In these cases, the cfa () function reported warning messages about convergence issues, but not about negative residual variances. These models were counted as nonconvergent cases, but not Heywood cases. We used the term Heywood cases exclusively for converged models with negative residual variances. For most of the Heywood cases, variable Y3 was the problem showing up in 47 out of the 51 Heywood cases with WLS estimators, 34 out of 34 with WLSMV and 33 out of 33 with ULS. In only 4 Heywood cases, with WLS estimators, the residual variances of Y2 were negative. Figure 1 shows the histogram of the 34 negative residual variances using the WLSMV estimator with delta parameterization. Most negative values are small. The range of negative values was similar for WLS and ULS, but the frequency of negative values was larger for WLS and about the same for ULS.

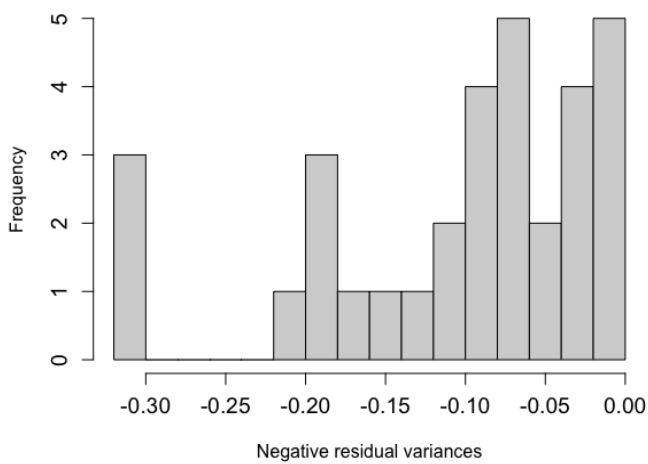

*Figure 1*. Histogram of negative residual variances obtained using the WLSMV estimator with delta parameterization.

*Theta parameterization*. Of the 100 datasets generated, nonconvergence was found for 54 datasets with WLS estimators, 34 datasets with WLSMV and 33 datasets with ULS, totaling 121 cases. Of the 121 cases of nonconvergence, the reported loadings were extremely large (positive or negative) for at least one predictor (always for Y3), and in 4 cases, extreme loadings were reported for 2 variables (Y2 and Y3). In a few rare cases, extreme loadings for variable Y3 were found in converged models: 2/46 with WLS, 2/66 with WLSMV, and 2/67 with ULS, and none was found for Y2. Finally, for all datasets with negative residual variances or nonconvergence when using delta parameterization, convergence issues were encountered when the data were fitted to theta parameterized models. This result illustrates that the negative residual variances in delta parameterized models cooccur with nonconvergent cases in theta parameterized models.

The WLS estimator reported the most problematic results in both delta and theta parameterized models compared to the other two limited information estimators, WLSMV and ULS, in line with the findings by Paek et al. (2018). The WLS estimator is known to be based on a complex algorithm that works best for extremely large datasets, which aligns with our findings. Our results also show that in comparison with WLS, both the WLSMV and the ULS estimators perform better and are more robust to possible negative variance problems in delta parameterized models and to convergence issues in theta parameterized models. It is therefore not surprising that WLSMV is the recommended estimator for ordinal factor models. WLSMV is more efficient than ULS although ULS also performs very well in other respects not investigated here (Yang-Wallentin & Joreskog, 2010).

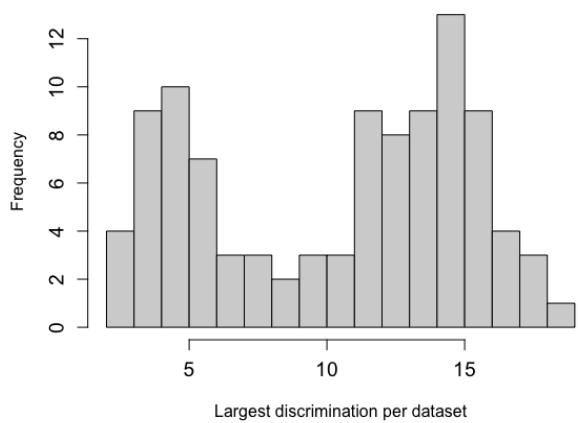

*Figure 2*. Histogram of the largest discrimination parameters per dataset.

*IRT model estimation*. All IRT models fitted to the generated data converged though some showed extremely large discrimination parameter estimates. Discrimination estimates of Y3 ranged from 2.26 to 18.77, and discrimination estimates of Y2 ranged from 1.24 to 13.89. For 58 out of 100 datasets, the estimated Y3 discrimination is larger than 10, and only for 1 other dataset, the estimated Y2 discrimination is larger than 10. None of the datasets showed both estimated discriminations larger than 10. Figure 2 shows a histogram of the most extreme discrimination parameter estimates per dataset. The discrimination estimates for the 34 Heywood cases obtained with delta parameterization and WLSMV ranged from 10.18 to 17.85; they are among the highest 58 values in Figure 2.

## Application-based Illustration

The data for the application are a subset from a larger dataset[2] (Yotebieng et al., 2016: Yotebieng, Fokong, & Yotebieng, 2017), concerning five PHQ9 items with N=200, randomly selected from a larger sample of respondents. The purpose of analyzing these data is to illustrate the differences among delta and theta parameterized ordinal factor models and IRT models with respect to the issues discussed earlier. The PHQ9 is a self-reported depression questionnaire intended to be one-dimensional (Spitzer, Kroenke, & Williams, 1999), and in this study, we included items 1, 2, 3, 7, and 8 referencing anhedonia, feeling depressed, sleep problems, concentration problems, and movement problems, respectively. Dichotomized responses were obtained by recoding 0 and 1 as 0, and 2 and 3 as 1. The data were then fitted to delta and theta parameterized one-factor ordinal factor models with the recommended WLSMV estimators using lavaan and to an IRT model with MML using ltm.

The results of the delta parameterized factor model are shown in Table 3. The model had a good fit, with RMSEA = 0.00 and SRMR = 0.044, but the residual variance of item 2 (feeling depressed) was negative, which means that a Heywood case was encountered.

---

[2] The data are from a study funded by grant R01HD075171 from NIHCD.

Table 3

Factor solution for the PHQ9 example

|        | Standardized Loading | Residual Variance |
|--------|----------------------|-------------------|
| Item 1 | 0.795                | 0.368             |
| Item 2 | 1.011                | -0.021            |
| Item 3 | 0.744                | 0.446             |
| Item 7 | 0.644                | 0.585             |
| Item 8 | 0.672                | 0.548             |

The theta parameterized model failed to converge, as expected. The loading of item 2 was reported as 14850.40 when the iterations were stopped. Similarly high loadings are usually found when delta parameterized models failed to converge, but for these data, the delta parameterized model did converge, albeit with a negative residual variance estimate. The application suggests that the same dataset showing Heywood case in the delta parameterized model showed nonconvergence in the theta parameterized model, for reasons explained earlier.

Finally, we fitted an IRT model to the same data using the ltm package in R. The results are shown in Table 4. The model converged, but the problematic item 2 was reported with a discrimination of 13.917, an extreme value on the logit scale.

Table 4

IRT estimates for PHQ9 example

|        | Difficulty | Discrimination |
|--------|------------|----------------|
| Item 1 | 0.463      | 2.369          |
| Item 2 | 0.609      | 13.917         |
| Item 3 | 0.541      | 1.903          |

| Item 7 | 1.647 | 1.540 |
| Item 8 | 1.933 | 1.752 |

The PHQ9 application illustrates how analyses of the same data can yield quite different outcomes depending on the approach, and yet are threaded by the same underlying problem. Using delta and theta parameterized factor models and IRT models, we showed three different ways the same problem is manifested in model estimation. To resolve the estimation issue in the delta and theta parameterized factor models for the present PHQ9 dataset, one can add correlated residuals for items 1 and 2 and for items 2 and 8 in the model specification. It is more complex to deal with remaining dependencies in an IRT model though local dependencies can be detected using well-functioning methods (e.g., Edwards, Houts, & Cai, 2018). How to extend an IRT model to accommodate local dependencies is not yet clear, but see Braeken, Tuerlinckx, and De Boeck (2007) for a copula approach, and Hoskens and De Boeck (1997) and Ip (2002) for other options.

## Discussion

The theoretical discussion and the illustrations show that the same problem appears in different forms depending on the method for analysis. Delta and theta parameterized ordinal factor models and IRT models produce different anomalies resulting from the same problem: a negative residual variance (in delta parameterized ordinal factor model), nonconvergence (in theta parameterized ordinal factor model) and largely outlying discrimination parameter estimates. Between the limited information approach and the full information approach, the full information approach is theoretically better. Therefore, when analyzing ordinal data, one may start with an IRT analysis (or an ordinal factor model using full information estimation, which is a roughly equivalent approach). If the results show an extremely large discrimination estimate, we recommend a further analysis. Because it is known that local dependencies can lead to large discrimination values (Edwards et al., 2018; Ip, 2000; Tuerlinckx & De Boeck, 2001), one may use local dependence statistics to investigate the problem. Alternatively, one may apply a delta parameterized ordinal factor model and verify whether a negative residual variance shows up, and if it does, one may investigate the residual correlations by fixing the residual variance as zero instead of allowing it to be freely estimated as negative.

If the theta parameterized ordinal factor model is the first choice, one would likely encounter nonconvergence. In that case, one can move to a delta parameterized model instead and investigate the residual correlations. There are other possible causes of nonconvergence, but a Heywood case is one of them, so that a delta parameterized analysis may aid a diagnosis.

An evident general recommendation is to work with a large sample size. We have worked with N=200. Larger is better, but to compare the three methods of analysis, the sample size is less important. If the true correlation matrix does not imply a Heywood case, a large sample size helps to avoid Heywood cases. However, if the true correlation matrix does imply a Heywood case, a large sample size would increase the probability of finding a Heywood case.

A limitation of our study is that it does not offer a more fundamental explanation for Heywood cases, and therefore a more detailed solution for the problem. Our purpose was to explain and report the different appearances of the same problem under the three different approaches. The causes of Heywood cases and nonconvergence issues (in theta parameterized models) are beyond the scope of our study. For causes and solutions of Heywood cases in linear factor models, we refer readers, among others, to Chen et al. (2001). A further investigation is required to find out how well the issues with linear factor models apply to models for ordered-category data, including binary data. Nonetheless, our comparative study contributes to the interpretation of outlying discrimination values encountered in IRT models and nonconvergence issues encountered in theta parameterized factor models.